\documentstyle[aps,prl,psfig]{revtex}

\newcommand{\be}{\begin{equation}}
\newcommand{\ee}{\end{equation}}

\newcommand{\bex}{\begin{eqnarray}}
\newcommand{\eex}{\end{eqnarray}}
\begin{document}

\title{Classical Communication via the Einstein-Podolsky-Rosen Channel Alone: A 
Proposed Experimental Test}
\author{R. Srikanth\thanks{e-mail: srik@iiap.ernet.in}}
\address{Indian Institute of Astrophysics, 
Koramangala, Bangalore- 34, Karnataka, India.}
\maketitle \date{}


\begin{abstract} 
We uncover an apparent instance of classical information transfer via only the 
Einstein-Podolsky-Rosen channel in a quantum optical protocol between Alice and 
Bob, involving two-photon maximal path entanglement and based on a recent
Innsbruck experiment. The signal is traced to the appearance of coherent 
reduction due to the onset of
spatial degeneracy in the eigenvalue spectrum for Alice's measurement. We 
present our result primarily as an issue for experimental testing rather than 
as a definitive prediction at this stage.
\end{abstract}

\parskip 3pt

~~~~~~~~

\section{Introduction}
Does quantum nonlocality \cite{epr}, now attested by a vast body of 
experimental evidence \cite{int,exp,sca00}, consist 
only in the local inaccesibility of information
or, in addition, a nonlocal transfer of information \cite{hen97}? The answer 
seems to lie subtly hidden in the quantum formalism.
The evidence for spacelike transfer of 
information remains circumstantial, as in quantum teleportation 
\cite{ben93} and remote state 
preparation \cite{ben00}-- where it can be shown {\em a posteriori} that the 
(in principle, infinite) classical information
about a quantum state is nonlocally transferred at the cost of only
2 bits \cite{sri0104081}-- or controversial, as in the violation of the
Bell inequalities \cite{bel64,exp}, which tells us 
that quantum mechanics (QM) cannot be local-realistic. 
However, it is usually agreed that quantum nonlocality is causal \cite{nosig}.
Here we present a test of the question raised at the beginning. 

The article is arranged as follows.
The experimental set-up for and the basic idea behind
the test are presented in the next section. 
The main result is derived more rigorously in Section \ref{main}.
This result in physically interpreted 
in Section \ref{degen}, where we show that it owes 
its origin to the onset of eigenvalue spectrum degeneracy in quantum 
measurement. We then conclude with a final brief section.
In view of the absence of a standard
method to handle quantum measurement with degenerate outcomes,
experimental tests would be valuable in confirming or refuting the 
effect predicted here to be possible.
Further careful scrutiny of the problem is needed also 
in view of the far-reaching implications of the question considered.

\section{Experiment}\label{xper}
\begin{figure}
\centerline{\psfig{file=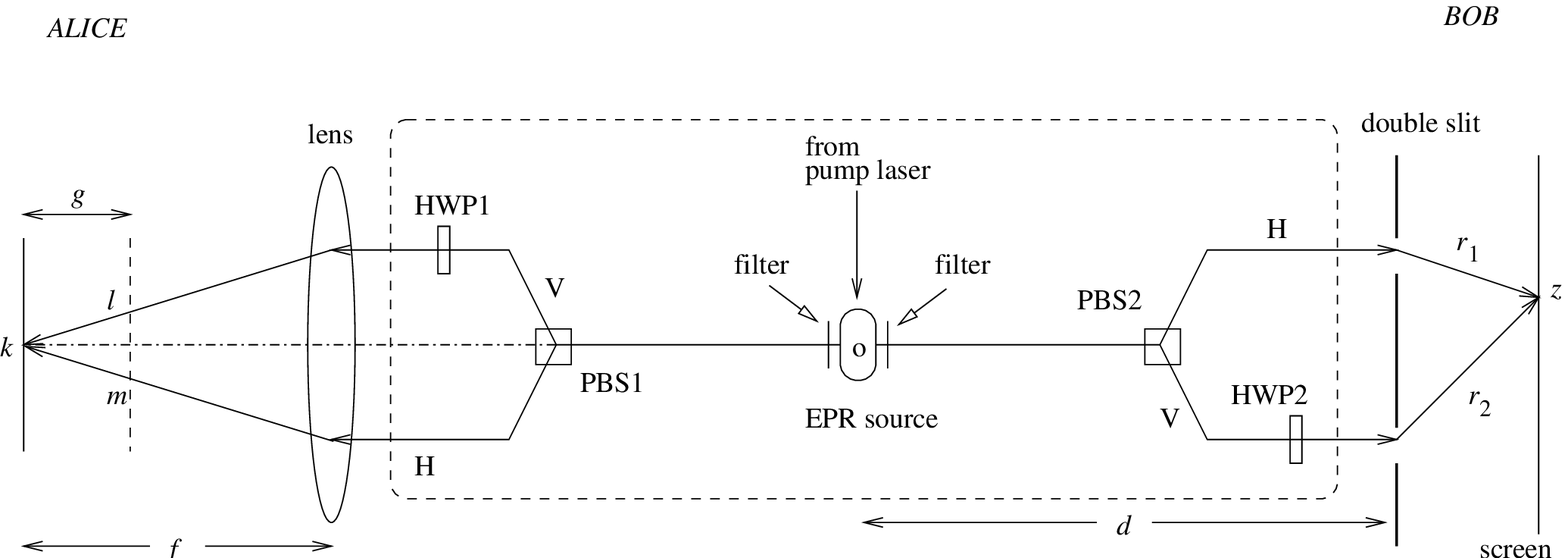,width=16.5cm}}
\vspace*{0.5cm}
\caption{Alice and Bob share a thin pencil of polarization-entangled biphotons 
in pure state, from an EPR source at $o$. By means of a pair of polarizing 
beam splitters (PBS1 and PBS2) and half-wave plates (HWP1 and HWP2), 
polarization entanglement is converted to path-entanglement. Depending on 
whether she observes her photon at the focus of her lens, or off the focal 
plane, she cannot or can obtain path information for Bob's photon, thereby 
permiting or prohibiting the latter's interference. The part of the experiment 
enclosed in the dashed box prepares the biphoton in used to prepare the 
biphoton in a path-entangled state.}
\label{yi}
\end{figure}

The proposed experimental test, based on ideas involved in a recent 
interesting quantum 
optical experiment performed by Zeilinger's group at Innsbruck \cite{zei00}, 
involves two observers,
usually called Alice and Bob, sharing path-entangled biphotons (pairs of 
entangled photons) from a suitable Einstein-Podolsky-Rosen source at point $o$.
One suggested candidate for the source is a nonlinear crystal (e.g., BBO) with 
appropriate optical
pre-processing, enclosed by the dashed box in Figure \ref{yi}, which is a 
`folded-out' schematic of the proposed experiment. 
The crystal produces polarization-entangled biphotons 
via spontaneous parametric 
down-conversion (SPDC) in type II phase matching \cite{kwi95,bit01}, pumped by
a suitable laser beam (eg., Argon laser with $\lambda$ = 395 nm).
A filter restricts outgoing photons to a small bandwidth 
about the downconverted frequency. 

The output from the crystal beyond the dashed box 
is the maximally entangled state
\be
\label{output}
|\psi\rangle_{AB} = \frac{1}{\sqrt{2}}(|HV\rangle_{AB} - |VH\rangle_{AB}),
\ee           
where $\{|V\rangle$, 
$|H\rangle\}$ represent vertical and horizontal polarization states
and subscripts $A$ and $B$ represent Alice's and Bob's photon.
On both sides of the crystal, there is: (1) a polarizing 
beam-splitter (PBS1 or PBS2) to seperate the beams in such a way that 
eventually the $H$ 
beam is parallel to the $V$ beam; (2) a half-wave-plate (HWP1 or HWP2) in the
$V$ paths to rotate them to $H$ in order to permit interference
(though the labels retain
their original polarization values for the sake of uniformity of notation). 
In this way, maximal polarization entanglement is converted to maximal
path entanglement. 

Alice is equipped with a lens of focal length $f$, and a movable detector 
system that can be positioned at distance $f$ or $f - g$ from the lens.
Bob is equipped with a Young's double-slit interferometer, located at distance 
$d$ from the plane containing the crystals. 
By classical optics, parallel rays entering
Alice's lens converge to a single point on the focal plane. Because of 
path entanglement, a
detection by Alice at some point on the focal plane
will leave Bob's photon in a superposition of parallel rays, i.e,
a definite `momentum state',
but with its point of origin in the source indeterminate, as expected
on basis of the Uncertainty principle. This has been confirmed 
by the interference pattern seen in Bob's photons detected in
coincidence with Alice's measurement at her focal plane \cite{zei00}. 
On the other hand, if Alice advances her detector system to a distance
closer to the lens, distinguishability is restored and no interference is
possible.

In the present protocol, Alice will choose to position her detector 
at distance $f$ or $f - g$ from the lens (Figure \ref{yi}). 
Alice's detector must be wide enough to intercept all $A$
photons. In practice, one detector element (the scanning tip of a fiber optic
element) is sufficient, to be positioned at $k$, for the 
focal plane measurement, or at either $l$ or $m$, for the off-focal-plane
measurement (since a non-detection is also a measurement). Suppose she 
positions her detector on the plane at distance $f - g$.  She will in effect
detect $A$ at point $l$ or $m$. Since her two possibile outcomes
are distinguishable, and because of entanglement, she correspondingly leaves 
Bob's photon in one of the states $|H\rangle$ or $|V\rangle$, respectively.
Neither detection by her produces an
interference in the coincidence counts at Bob's interferometer because
she has acquired path information for Bob's photon. 
Therefore, Bob will find no interference pattern on his
screen in his single counts.

On the other hand, suppose she positions it at the focus $k$ of the lens. 
She will detect a click at point
$k$. Since she cannot distinguish whether her detection was generated by a 
photon coming through the upper or lower path, by Feynman's dictum, both paths 
interfere at $k$, and, because of path entanglement, correspondingly she is 
expected to leave Bob's photon in the superposition state
\be
\label{M}
|M\rangle_B \equiv \frac{1}{\sqrt{2}}(\alpha |H\rangle_B + \beta |V\rangle_B)
\hspace{1.0cm}(|\alpha|^2 + |\beta|^2 = 1),
\ee
where $\alpha$ and $\beta$ are path dependent phase factors
(cf. Section \ref{degen}). Therefore, in this case, Bob
will find a Young's double slit pattern in the coincidence counts, produced by 
the interfering $H$ and $V$ rays, rotated by the half-wave plate HWP2. 
Further, because the coincidence counts involve only one possible outcome for
Alice, he in fact finds the interference pattern in his singles counts. 

What evidence do we have that Bob's photon is indeed
projected into the state Eq. (\ref{M}) by Alice's focal
plane measurement? One way to understand it is to view Alice's photon $A$ as a
handle on Bob's photon's path. Registering $A$ at $k$, irreversibly
destroys path information about Bob's photon, so the two $B$ paths are
free to interfere, but otherwise, not. 
This complementaristic interpretation is well borne out by
implementations of the delayed choice experiment \cite{her95}. Furthermore, 
coincidence
measurements on two-particle interference experiments \cite{int} clearly
show that the point where one of the entangled pair is registered
acts like a virtual source for the origin of the modes whose interference
determines the probability distribution for the localization of its
twin photon. 
A further corroboration of this reasoning, closer to the
experiment at hand, comes
from the Innsbruck experiment \cite{zei00}, where it is found that 
when the signal beam is focussed to a detector using a lens, an interference
pattern is detected in the idler coincidence counts, as discussed earlier. 

Finally, we want to
note that visualizing state $|M\rangle_B$ as the projection that results from
Alice's measurement is a sort of concession to our quantum mechanical 
intuition, but is not necessary for the calculation of probabilities in
the quantum optical formalism, as shown in the following section.
Bob discerns whether Alice measured at the focal plane or not
depending on whether or not he finds the 
tell-tale interference pattern. Clearly, this classical signal can be 
transmitted arbitrarily fast by increasing $f$ and/or $d$ indefinitely. 
This completes the basic idea of the proposed nonlocal classical signaling 
test, which will be examined critically in the next two sections. 

\section{Derivation}\label{main}

We now derive quantitatively the result given in the preceding section.
The four-mode state vector of the SPDC field incident on Alice's and Bob's
detectors is given by:
\be
\label{spdc}
|\Psi\rangle = |{\rm vac}\rangle + \epsilon(|hv\rangle - |vh\rangle)
\ee
where $|{\rm vac}\rangle$ is the vacuum state, 
$|h\rangle$ and $|v\rangle$ are the Fock state modes propagating along the
$H$ and $V$ arms of the experiment and $\epsilon (\ll 1)$ depends on the
pump field strength and the crystal nonlinearity. 
The positive frequency part of the electric field at an arbitrary point $z$ 
on Bob's screen is:
\be
\label{bobfjeld}
E_z^{(+)} = e^{ikr_D}\left(e^{ikr_1}\hat{h} + e^{ikr_2}\hat{v}\right),
\ee
where $\hat{h}$ and $\hat{v}$ are the annihilation operators for the
$h$ and $v$ modes, respectively. $r_D$ is the distance from the source
$o$ via PBS2 to the upper/lower slit on Bob's double slit diaphragm;
$r_1$ ($r_2$) is the distance from the upper (lower) slit to $z$ (Figure
\ref{yi}).

If Alice positions her dectector at point $l$ or $m$ on the plane at
distance $f - g$ from the lens, the positive frequency part of the electric 
field at point $l$ or $m$ is
\bex
\label{alicefjeldb}
E_l^{(+)} = e^{ikr_L}\hat{v}, \hspace{1.0cm}
E_m^{(+)} = e^{ikr_M}\hat{h},
\eex
where $r_L$ ($r_M$) is the distance from 
$o$ via PBS1 along the upper (lower) path through the lens upto point $l$
($m$). For simplicity, we set $r_L = r_M$. Now,
if Alice positions her dectector at the focal plane, the positive frequency 
part of the electric field at point $k$ is
\be
\label{alicefjelda}
E_k^{(+)} = e^{ikr_K}\left(\hat{h} + \hat{v}\right),
\ee
where $r_K$ is the distance from 
$o$ via PBS1 along the upper or lower path through the lens upto point $k$.
Again, for simplicity, the distances along the two paths have been taken to be
identical. 

The coincidence count rate $R$ for simultaneous measurements by Alice and
Bob is given by the absolute square of the second order correlation function
$\langle\Psi|E^{(+)}_yE^{(+)}_z|\Psi\rangle$ ($y = k, l, m$). This
is proportional to the probability for Alice's and Bob's correlated 
measurements. If Alice positions her dectector at point $l$ or $m$ on the 
plane at distance $f - g$ from the lens, the coincidence rate 
for detections by Alice and by Bob at $z$ is
\be
\label{Rg}
R_g \propto |\langle\Psi| E_y^{(+)}E_z^{(+)}|\Psi\rangle|^2 
    \propto \epsilon^2 ,
\ee
where $(y = l, m)$ and we have used Eqs. (\ref{spdc}), (\ref{bobfjeld}) and 
(\ref{alicefjelda}). As the coincidence rates for Alice's both detections
are uniform, Bob finds a uniform intensity pattern on his screen. 
On the other hand, if she positions her detector at focus $k$,
the coincidence rate is given by 
\be
\label{Rf}
R_f \propto |\langle\Psi| E_k^{(+)}E_z^{(+)}|\Psi\rangle|^2 = 
    \epsilon^2\{1 + \cos(k\cdot[r_1 - r_2])\},
\ee
which is equivalent to a conventional Young's double slit pattern. Because of
the focussing, no other coincidence terms are involved. So the interference
pattern Eq. (\ref{Rf}) is in fact seen in Bob's singles counts.
In an actual implementation, Eqs. (\ref{Rg}) and 
(\ref{Rf}) must be further modified to take
into consideration the single slit diffraction pattern and the profile of the
down-converted laser beam.

Usually, in biphoton interference \cite{int} experiments,
Alice's photon is not focussed to a single point, but allowed to
spread out according to the beam profile or through a single/double
slit system. Therefore
Bob would see an interference pattern of the type Eq. (\ref{Rf}) 
averaged over various Alice's detection positions, which
smears Bob's pattern to a uniform distribution that is indistinguishable
from that in Eq. (\ref{Rg}). In the above experiment,
by the focussing of Alice's beam, this smearing is-- crucially-- checked.

Are there some other reasons that come into play
that somehow restore the distinguishability between the two paths in Alice's
focal plane measurement? Ultimately, only an experimental test can adjudicate. 
Suppose Bob's beam diverges slowly, i.e,
$s\theta \ll \lambda$, where $s$, $\theta$
and $\lambda$ are slit-width, divergence angle and wavelength, respectively
An unentangled laser beam satisfying this condition will produce a Young's
double slit pattern. After Alice's focal plane measurement, 
Bob's beam is indeed disentangled into such an unentangled laser beam. 
Of course, the wavefront
of Bob's disentangled beam will be modified because only annular regions about
the lens's principal axis have the same phase (since they have the same
$r_K$). Nevertheless, provided the
double slit is positioned symmetrically about the lens's axis, and Bob's
downconverted (entangled) beam satisfies the usual interference criterion
$s\theta \ll \lambda$, then a possible interesting outcome for the above
experiment can be expected.

We note that unlike the case with position-momentum entanglement, a spreading
of polarization-entangled light of laser is not required by the
Uncertainty principle. Therefore, in principle, polarization-entangled thin
pencils satisfying the above slow divergence condition are easier to prepare 
than position-momentum entangled beams. This is 
the reason why the present experiment is simpler to implement than
that presented in Ref. \cite{sri2223}, in which considerable optical 
preprocessing is needed to bring about a position-momentum entangled equivalent
of the above experiment, and which, incidentally,
is closer to the Innsbruck experiment \cite{zei00}. In this regard, we note 
that that no sudden spreading of
$B$ occurs on account of $A$'s localization even if the input is light
position-momentum entangled \cite{kim}. 

Two suggestions for the preparation of path-entangled pencils: (1) blocking 
out photons $A$ and $B$ with a shield, except at two small opposing holes, 
one on each shield, from which fiber optic cables of equal
length lead to the respective polarizing beam splitter; (2) 
selective Bell state measurement using appropriate linear optics 
on two seperable thin laser beams (in this connection, cf. Ref. \cite{lam01}).

\section{Quantum mechanical picture}\label{degen} 

Although interference experiments rightly belong to the domain of quantum
optics (QO), many of them can usually be translated into quantum mechanical 
language (for example, cf. Ref. \cite{lou} as regards the Mach-Zehnder 
interferometer and Ref. \cite{sri154} as regards the delayed choice experiment 
\cite{whe94,hom94}). Sometimes the
latter version can be easier to physically interpret. 
Reverting back to the QM notation 
of Section \ref{xper},  we find the reduced density matrix 
for Bob's photon if Alice measures in the `focal plane'
basis, namely $(\rho_f)_B$, and that if she measures off the focal plane, 
namely $(\rho_g)_B$, to be
\bex
(\rho_f)_B &=& |M\rangle_B\langle M|_B = \frac{1}{2}(\alpha|H\rangle_B + 
  \beta|V\rangle_B)(\alpha^*\langle H|_B + \beta^*\langle V|_B) \nonumber \\
(\rho_g)_B &=& \frac{1}{2}\left(|H\rangle_B\langle H|_B +
                    |V\rangle_B\langle V|_B\right).
\eex
The classicality of the signal is an expression of the fact that 
$(\rho_f)_B \ne (\rho_g)_B$.

In order to trace the origin of the classical signal, let us interpret the
results of the preceding section in the QM Schr\"odinger picture by inserting
the appropriate phase factors.
On each path, 
\be
\label{transA}
|H\rangle_X \longrightarrow e^{ik\cdot x_{HX}}|H\rangle_X;
\hspace{1.0cm}
|V\rangle_X \longrightarrow e^{ik\cdot x_{VX}}|V\rangle_X,
\ee
where $x_{HX}$ ($x_{VX}$) is the distance along the $H$ ($V$) path on beam $X$
($X = A, B$) from the source at $o$. Beyond the double slit, $B$ is
transformed into Bob's screen measurement basis according to:
\be
\label{transB}
|H\rangle_B \longrightarrow e^{ik\cdot r_1(z)}|z\rangle
\hspace{1.0cm}
|V\rangle_B \longrightarrow e^{ik\cdot r_2(z)}|z\rangle ,
\ee
where $|z\rangle$ is the eigenstate corresponding to $B$ being found at $z$, 
an arbitrary detector element. Bob's measurement that localizes his particle 
at $z$ is given by the usual von Neumann projector, 
$\hat{P}_z \equiv |z\rangle\langle z|$. 

After inserting the spatial dependences Eqs. (\ref{transA}) and (\ref{transB})
into the biphoton state vector Eq. (\ref{output}), we have
\be
\label{output0}
 |\psi\rangle_{AB} \longrightarrow \sum_z \left(|Hz\rangle_{AB}
 e^{ik[r_A + r_D + r_2(z)]} - |Vz\rangle_{AB}e^{ik[r_A + r_D + r_1(z)]}\right),
\ee
with $r_A$ set to $r_L = r_M$, for off-focal-plane measurement, or to $r_K$,
for focal plane measurement.

These two measurement planes are equivalent
to two different observables. The spectral decomposition for the observable 
corresponding to measurement on the plane at distance $f - g$ from the lens 
can be written as
\be
\label{Og}
\hat{O}_g = l|V\rangle_A\langle V|_A + m|H\rangle_A\langle H|_A.
\ee
According to the von Neumann projection postulate \cite{vN}, the probability to
find $m$ or $n$ is given by the expectation value of the corresponding projector
\be
\label{Ebc}
\hat{P}_l \equiv |V\rangle_A\langle V|_A, \hspace{1.0cm}
\hat{P}_m \equiv |H\rangle_A\langle H|_A
\ee
in the state $|\Psi\rangle_{AB}$ in Eq. (\ref{output0}) (setting $r_K$ to
$r_L = r_M$) \cite{vN}.

The observable corresponding to measurement on the focal plane 
is seen to be degenerate
in space, i.e., both eigenstates $|H\rangle$ and $|V\rangle$ have the
same position eigenvalue $k$. Hence, in analogy with Eq. (\ref{Og}), we write:
\be
\label{Of}
\hat{O}_f = k|V\rangle_A\langle V|_A + k|H\rangle_A\langle H|_A.
\ee
(An `energetic' analogy would be switching a magnetic field on or off to render 
the spin eigenstates of an entangled electron non-degenerate or degenerate in 
a local energy measurement.) 
A measurement of position yields the value $k$, but what is the corresponding
projector? Clearly the von Neumann projection postulate cannot
handle this case and must somehow be extended. The problem of degenerate
measurement was first considered by L\"uders \cite{lud51} and has been the
subject of recent renewed interest \cite{bro01}. 

Two related problems here are that of: (a)
calculating the probability for obtaining the eigenvalue corresponding to the
degenerate subspace and, (b) determining the state in which the system is left
if the degenerate eigenvalue is found. 
The latter problem for QO is necessarily
different from that in QM because  no particle annihilation figures in QM,
unlike in QO. Hence, a QM interpretation is applicable 
only with respect to the former problem. In L\"uders' \cite{lud51}
formalism for extending the projection postulate, the projector $\hat{P}_k$ 
for Alice's focal plane measurement is given by $\hat{P}_k = \hat{P}_l + 
\hat{P}_m$. If valid, this extension
would indeed be sufficient to prohibit the classical signaling.
However, one can verify that $\langle (\hat{P}_l + \hat{P}_m) \otimes 
\hat{P}_z\rangle$ does not reproduce the
interference Eq. (\ref{Rf}), where $\langle\cdots\rangle$ represents
expectation value with respect to the state Eq. 
(\ref{output0}) setting $r_A = r_K$. The reason is that
it does not permit crosstalk between Bob's $H$ and $V$ modes, needed to
explain Bob's coincident interference pattern in Eq. (\ref{Rf}), and
more generally, the interference seen in coincidence with focal plane 
detections of $A$ in the Innsbruck experiment \cite{zei00}. 

It turns out that the form of
the degenerate projector that agrees with Eq. (\ref{Rf}), and thus 
rightly represents $A$'s electric field in Eq. (\ref{alicefjelda}), is
\be
\label{Ea}
\hat{P}_k \equiv \hat{P}_{l+m} = 
(|H\rangle_A + |V\rangle_A)(\langle H|_A + \langle V|_A)
\ne \hat{P}_l + \hat{P}_m, 
\ee
in view of Eq. (\ref{Ebc}).  By direct computation,
one can verify that $\langle \hat{P}_{l+m} \otimes \hat{P}_z\rangle$ 
indeed reproduces the interference Eq. (\ref{Rf}).
This implies that {\em in projecting a state vector to a state 
corresponding to a degenerate eigenvalue, the amplitudes of the eigenstates 
in the degenerate subspace superpose}. In other words, the probability 
amplitude that a measurement finds a degenerate 
eigenvalue is given by the sum of the degenerate amplitudes. 
The consequences of this for state representation and state vector reduction,
both in seperable and entangled systems, and experimental tests of whether the 
validity of such a `coherent projection' and of the associated `coherent 
reduction' can be extended beyond QO to the case of energy degeneracy in QM 
proper, are taken up in the future. Coherent reduction brings further richness 
to the essential quantum phenomenon of superposition. What is encouraging is 
that quantum optical tests for it are well feasible (in a related vein, cf.
Ref. \cite{ghi98}).
 
According to the foregoing analysis, the classical nonlocal signal in the 
proposed experiment owes its origin to the fact that Alice can choose to make 
either a von Neumann measurement
in the non-degenerate off-focal-plane basis, or a degenerate measurement in
the focal plane basis, thereby disentangling Bob's photon by projecting it 
completely or coherently. As a result, it requires only the EPR
channel and no additional classical communication channel.
This is not incompatible with the conclusion of
Refs. \cite{nosig}, where implicitly only complete von Neumann measurements in 
a non-degenerate basis are considered. 

The new result derived here is in fact implicit in the
Innsbruck experiment \cite{zei00}, and we would not expect its possible 
positive outcome if the Innsbruck experiment would not have found 
interferences in Bob's coincidence counts for Alice's
focal plane measurement. And yet this latter counterfactual hypothesis would 
not be possible without the abandonment of the usual quantum optical formalism 
for calculating interferences at second order. Therefore, the classicality of
the signal in a sense lurks in the familiar double slit interference.

\section{Conclusion}

Feynman noted that the central mystery of QM--
namely, superposition-- is encapsulated by the double-slit interference 
\cite{fey65}. Coherent reduction, as discussed above,
 adds a further perspective to this `mystery'. 
 Technically speaking, an (improbable) positive outcome of the
experiment is not incompatible with QM itself, since
the features of QM that guarantee relativistic causality-- namely,
linearity \cite{gis90,rig95}, unitarity \cite{woo82}
and the tensor product character of the Hilbert space of composite quantum
systems \cite{pea00}-- are essentially non-relativistic. On the other hand, a 
null result for the proposed experiment, although less obviously explained, 
would be easier to accept.


\newpage


\end{document}